\documentclass[conference, 10pt]{IEEEtran}

\usepackage{multicol}
\usepackage{amsmath,epsfig}
\usepackage{amssymb}
\usepackage{booktabs}
\usepackage{amsthm}
\usepackage{graphicx}
\usepackage{caption}
\usepackage{subcaption}
\usepackage{rotating}
\usepackage{floatflt} 
\usepackage{paralist} 
\usepackage{algorithm} 
\usepackage{xcolor}
\usepackage{color}
\usepackage{epstopdf}
\usepackage[normalem]{ulem}
\usepackage{float}
\usepackage{todonotes}
\usepackage{comment}

\newcommand{\mypar}[1]{{\bf #1.}}
\theoremstyle{definition}
\newtheorem{defn}{Definition}
\newtheorem{myLem}{Lemma}
\newtheorem{myAlg}{Algorithm}
\newtheorem{myThm}{Theorem}
\newtheorem{myCorollary}{Corollary}
\newcommand{\R}{\ensuremath{\mathbb{R}}}

\DeclareMathOperator{\Id}{I}
\newcommand{\argmin}{\mathbf{arg min}}

\DeclareMathOperator{\BL}{BL}
\DeclareMathOperator{\BLT}{ABL}

\DeclareMathOperator{\GS}{GS}

\def\x{\mathbf{x}}
\def\h{\mathbf{h}}

\def\xhat{\widehat{\x}}

\def\y{\mathbf{y}}

\def\u{\mathbf{u}}
\def\vv{\mathbf{v}}

\def\N{\mathcal{N}}

\def\V{\mathcal{V}}
\def\M{\mathcal{M}}

\DeclareMathOperator{\Adj}{A}

\DeclareMathOperator{\Um}{U}
\DeclareMathOperator{\Vm}{V}

\newcommand{\highlightChange}{\color{red}}
\def\HC{\highlightChange}

\begin{document}

\begin{multicols}{2}

\title{\vspace*{-3mm}Signal Recovery on Graphs:\\ Random versus Experimentally Designed Sampling\vspace*{-3mm}}
\author{
    \IEEEauthorblockN{Siheng Chen}
    \IEEEauthorblockA{ ECE\\
											Carnegie Mellon University\\
											sihengc@andrew.cmu.edu\\}
  \and
    \IEEEauthorblockN{Rohan Varma}
    \IEEEauthorblockA{ ECE\\
											Carnegie Mellon University\\
											rohanv@andrew.cmu.edu\\}
	\and
    \IEEEauthorblockN{Aarti Singh}
    \IEEEauthorblockA{ Machine learning\\
											Carnegie Mellon University\\
											aarti@cs.cmu.edu}
	\and
    \IEEEauthorblockN{Jelena Kova\v{c}evi\'c}
    \IEEEauthorblockA{ ECE \&  BME\\
											Carnegie Mellon University\\
											jelenak@cmu.edu}}										
\maketitle
\end{multicols}

\begin{abstract}
We study signal recovery on graphs based on two sampling strategies: random sampling and experimentally designed sampling. We propose a new class of smooth graph signals, called approximately bandlimited. We then propose two recovery strategies based on random sampling and experimentally designed sampling. The proposed recovery strategy based on experimentally designed sampling uses sampling scores, which is similar to the leverage scores used in the matrix approximation. We show that while both strategies are unbiased estimators for the low-frequency components, the convergence rate of experimentally designed sampling is much faster than that of random sampling when a graph is irregular\footnote{Due to lack of space, the proofs of lemmas and theorems are omitted and will be included in an expanded version of the paper.}. We validate the proposed recovery strategies on three specific graphs: a ring graph, an Erd\H{o}s-R\'enyi  graph, and a star graph. The simulation results  support the theoretical~analysis.
\end{abstract}

  \vspace{-3mm}
\section{Introduction}
We consider sampling and recovery within the framework of signal processing on graphs, which studies signals with an underlying complex structure~\cite{ShumanNFOV:13,SandryhailaM:14}. The framework models that underlying structure by a graph and signals by graph signals, generalizing concepts and tools from classical discrete signal processing.  The task of sampling and recovery is one of the most critical topics in the signal processing community. As the bridge connecting sequences and functions, classical sampling theory shows that a bandlimited function can be perfectly recovered from its sampled sequence if the sampling rate is high enough. The interest in sampling and recovery of graph signals has increased in the last few years~\cite{Pesenson:08, AnisGO:14, WangLG:14}.  In~\cite{ChenSMK:14}, authors proposed an algorithm to recover graph signals that have small variation based on random sampling. In~\cite{AnisGO:14, ChenVSK:15}, authors proposed a sampling theory for graph signals and show perfect recovery for bandlimited graph signals based on experimentally designed sampling.

In this paper, we propose a new class of graph signals, called~\emph{approximately bandlimited}. We then propose two recovery strategies based on random sampling and experimentally designed sampling, and bound the recovery error for the class of approximately bandlimited graph signals. We show that the proposed recovery strategies are unbiased estimators for low-frequency components and that experimentally designed sampling outperforms random sampling in terms of the convergence rate when a graph is irregular. We validate both recovery strategies on three specific graphs: a ring graph, an Erd\H{o}s-R\'enyi  graph, and a star graph. The simulation results support the theoretical analysis.

\vspace{-1mm}
\section{Discrete Signal Processing on Graphs}
We now briefly review discrete signal processing on graphs~\cite{SandryhailaM:14}, which lays a foundation for the proposed work.
We consider a  graph $G = (\V,\Adj)$, where $\V = \{v_0,\ldots, v_{N-1}\}$ is the set of nodes
and $\Adj \in \R^{N \times N}$ is the graph shift, or a weighted adjacency matrix. As the most basic filter defined on this graph, the graph shift represents the connections of the graph $G$, which can be
either directed or undirected. The
edge weight $\Adj_{n,m}$ between nodes $v_n$ and $v_m$ is a quantitative expression of the underlying relation between the $n$th and the $m$th node, such as a similarity, a dependency, or a communication
pattern. To guarantee that the filtered signal is properly scaled for comparison in the original one~\cite{SandryhailaM:14}, we normalize the graph shift, such that $|\lambda_{\max} (\Adj)| = 1$. Once the node order is fixed, the graph
signal can be written as a vector, $  \x \ = \ \begin{bmatrix}
 x_0, x_1, \ldots, x_{N-1}
\end{bmatrix}^T \in \R^N.$ The Jordan decomposition of $\Adj$ is~\cite{SandryhailaM:14}
\vspace{-1mm}
\begin{equation}
  \label{eq:eigendecomposition}
  \Adj=\Vm \Lambda \Um,
  \vspace{-1mm}
\end{equation} 
where the generalized eigenvectors of $\Adj$ form the columns of matrix $\Vm$ , $\Um = \Vm^{-1}$ (the norm of each row of $\Um$ is normalized to one), and
$\Lambda\in\R^{N\times N}$ is the block diagonal matrix of corresponding
eigenvalues $\lambda_0, \, \ldots \, \lambda_{N-1}$ of $\Adj$ ($1 = \lambda_0 \geq  \lambda_1 \geq \ldots, \,  \geq \lambda_{N-1} \geq -1)$. The~\emph{graph Fourier transform} of $\x \in \R^N$ is
  \vspace{-1mm}
\begin{equation}
  \label{eq:graph_FT}
  \widehat{\x} = \Um \x.
    \vspace{-1mm}
\end{equation}   The~\emph{inverse graph Fourier transform} is
$
 \x  =  \Vm  \widehat{\x}  = \sum_{k=0}^{N-1}  \widehat{x}_k \vv_k,
$
where $\vv_k$ is the $k$th column of $\Vm$ and $\widehat{x}_k $ is the $k$th component in $\widehat{\x}$. The vector $\widehat{\x}$ in~\eqref{eq:graph_FT} represents the
signal's expansion in the eigenvector basis and describes the frequency
components of the graph signal $\x$. The inverse graph Fourier transform
reconstructs the graph signal by combining
graph frequency components. When $\Adj$ represents an undirected graph, we have $\Um = \Vm^T$, and both $\Um$ and $\Vm$ are orthonormal. In general,  $\Vm$ may not be orthonormal; to restrict its behavior, we assume that
\begin{equation}
\label{eq:GFTB}
\alpha_1  \left\| \x  \right\|_2^2\leq  \left\| \Vm \x \right\|^2 \leq  \alpha_2 \left\| \x  \right\|_2^2, ~~{\rm for~all}~\x \in \R^N,
\end{equation}
where $\alpha_1, \alpha_2 > 0$, that is, $\Vm$ is a Riesz basis with stability constants $\alpha_1, \alpha_2$~\cite{VetterliKG:12}.  The eigenvalues  $\lambda_0, \, \ldots \, \lambda_{N-1}$ of $\Adj$, represent frequencies on the graph~\cite{SandryhailaM:14}. 

\section{Problem Formulation}
\label{sec:formulation}
We now review two standard classes of graph signals, and propose a new one, which connects the first two. We next describe the sampling and recovery strategies. In this way, we show the connection between this work and the previous work: graph signal inpainting and sampling theory on~graphs.

\subsection{Graph Signal Model}
\label{sec:model}
We focus on smooth graph signals, that is, the signal coefficient at each node is close to the signal coefficients of its neighbors. In literature~\cite{ChenSMK:14,ChenVSK:15}, two classes of graph signals have been introduced to measure the smoothness on graphs.
\begin{defn}
  \label{df:GS}
A graph signal $\x \in \R^N$ is~\emph{globally smooth} on a graph $\Adj \in \R^{N \times N}$ with parameter $\eta \geq 0 $,  when 
  \vspace{-1mm}
\begin{equation}
 \label{eq:GS}
\left\|  \x - \Adj \x \right\|_2^2 \ \leq \ \eta \left\| \x \right\|_2^2.
\end{equation}
  \vspace{-1mm}
Denote this class of graph signals by $\GS_{\Adj}( \eta)$.
\end{defn}
Since we normalized the graph shift such that $|\lambda_{\max} (\Adj)| = 1$; when $\eta \geq 4$, all graph signals satisfy~\eqref{eq:GS}. While the recovery of globally smooth graph signals has been studied in~\cite{ChenSMK:14} (leading to graph signal inpainting), global smoothness is a general requirement, making it hard to provide further theoretical insight~\cite{SharpnackS10}.
\begin{defn}
  \label{df:BL}
  A graph signal $\x \in \R^N$ is~\emph{bandlimited} on a graph $\Adj$ with parameter $K \in
  \{0, 1, \cdots, N-1\}$, when the graph frequency components
  $\widehat{\x}$ satisfies
  \vspace{-1mm}
  \begin{displaymath}
  \widehat{x}_k  \ = \ 0 \quad {\rm for~all~}  \quad k \geq K.
\end{displaymath}
  \vspace{-2mm}
Denote this class of graph signals by $\BL_{\Adj}(K)$.
\end{defn}
Note tht the original definition requires $\widehat{\x}$ be $K$-sparse, which unnecessarily promotes smoothness~\cite{ChenVSK:15}.  While the recovery of bandlimited graph signals has been studied in~\cite{ChenVSK:15} (leading to sampling theory on graphs), the bandlimited requirement is a restrictive requirement, making it hard to use in the real world applications. We thus propose a third class that relaxes it, but still promotes  smoothness. 
\begin{defn}
  \label{df:BLT}
  A graph signal $\x \in \R^N$ is~\emph{approximately bandlimited} on a graph $\Adj$ with parameters $\beta \geq 1$ and $\mu \geq 0$, when there exists a $K \in
  \{0, 1, \cdots, N-1\}$ such that its graph Fourier transform
  $\widehat{\x}$ satisfies
    \vspace{-2mm}
  \begin{equation*}
  \label{eq:BLT}
  \sum_{k = K}^{N-1} (1+k^{2\beta}) \widehat{x}_k^2 \ \leq \ \mu \left\| \x \right\|_2^2.
  \end{equation*}
    \vspace{-2mm}
Denote this class of graph signals by $\BLT_{\Adj}(K, \beta, \mu)$.
\end{defn}

We see that $\BL_{\Adj}(K)$ is a subset of $\BLT_{\Adj}(K, \beta, \mu)$ with $\mu=0, \beta=0$. The approximately bandlimited class allows for a tail after the first $K$ frequency components. The parameter $\mu$ controls the shape of the tail; the smaller the $\mu$, the smaller the energy contribution from the high-frequency components. The parameter  $\beta$ controls the speed of  energy decaying; the  larger the $\beta$, the larger the penalty on the high-frequency components. The class of $\BLT_{\Adj}(K)$ is similar to the ellipsoid constraints in~\cite{Johnstone:94}, where all the graph frequency components are considered in the constraints; thus, $\BLT_{\Adj}(K)$ provides more flexibility for the low-frequency components.

The following theorem shows the relationship between $\BLT_{\Adj}(K, \beta, \mu)$ and $\GS_{\Adj}(\eta)$.
\begin{myThm} 
\label{thm:BLKvsGS}
$\BLT_{\Adj}(K, \beta, \mu)$ is a subset of $\GS_{\Adj}(\eta)$, when 
\begin{equation*}
\eta \geq  \left( 1-\lambda_{K-1}+ \sqrt{    \frac{ 4\alpha_2 \mu} { (1+K^{2\beta})}  } \right)^2;
\end{equation*}   $\GS_{\Adj}(\eta)$ is a subset of $\BLT_{\Adj}(K, \beta, \mu)$, when 
\begin{equation*}
\mu \geq \frac{ 1+(N-1)^{2\beta}}{(1-\lambda_K) \alpha_1} \eta.
\end{equation*}
\end{myThm}  From Theorem~\ref{thm:BLKvsGS}, we see that when choosing proper parameters, $\GS_{\Adj}(\eta)$ is a subset of $\BLT_{\Adj}(K, \beta, \mu)$.
 \vspace{-1mm}
\subsection{Sampling \& Recovery}
We consider the procedure of sampling and recovery as follows: we sample $M$ coefficients in a graph signal $\x \in \R^N$ with noise to produce a noisy sampled signal $\y \in  \R^{M}$, that is,
 \vspace{-3mm}
\begin{eqnarray}
\label{eq:sampling}
\y  & = & \Psi \x + \epsilon \ \equiv \  \x_\M + \epsilon,
\end{eqnarray}
where $\epsilon \sim \N( 0, \sigma^2 \Id_{M \times M})$, $\M = (\M_0, \cdots, \M_{M-1})$ denotes the sequence of sampled indices, called~\emph{sampling set}, with $\M_i \in \{0,1, \cdots, N-1\}$ and $|\M| = M$, $\x_{\M}$ is the noiseless sampled coefficients, and the sampling operator $\Psi$ is a linear mapping from
$\R^N$ to $\R^M$,
 \vspace{-3mm}
\begin{equation}
\label{eq:Psi}
 \Psi_{i,j} = 
  \left\{ 
    \begin{array}{rl}
      1, & j = \M_i;\\
      0, & \mbox{otherwise}.
  \end{array} \right.
\end{equation}
 We then interpolate $\y$ to get $\x' \in \R^N$, which recovers $\x$ either exactly or approximately.  We consider two sampling strategies:~\emph{random sampling} means that sample indices are chosen from from $\{0, 1, \cdots, N-1\}$ independently and randomly; and~\emph{experimentally design sampling} means that sample indices can be chosen beforehand. It is clear that random sampling is a subset of experimentally design~sampling.

\section{Recovery Strategy}
\label{sec:upper}
We now propose two recovery strategies based on random sampling and experimentally designed sampling. In Section~\ref{sec:model}, we showed that a graph signal is smooth when its energy is mainly concentrated in the low-frequency components. For example, for the class $\BL_{\Adj}(K)$, all the energy is concentrated in the first $K$ frequency components and the graph signal can be perfectly recovered by using those first $K$ frequency components. The recovery strategies we propose here follow this intuition, by providing unbiased estimators for the low-frequency components.

\subsection{Recovery Strategy based on Random Sampling}
We consider the following recovery strategy.

\begin{myAlg}
\label{alg:ss1}
We sample a graph signal $|\M|$ times. Each time, we choose a node $i$ independently and randomly, and take a measurement $y_i$. We then recover the original graph signal  by using the following two steps:
\begin{eqnarray*}
  \widehat{x}_k^* & = & \frac{N}{ |\M|}\sum_{i \in \M} \Um_{ki} y_i,
  \\
  x_i^* & = & \sum_{k <  \kappa} \Vm_{ik} \widehat{x}_k^*,
\end{eqnarray*}
where $ x_i^*$ is the $i$th component of the recovered graph signal~$\x^*$.
\end{myAlg}

Algorithm~\ref{alg:ss1} aims to estimate the first $\kappa$ frequency components, and reconstruct the original graph signal based on these graph frequency components. The only tuning parameter in Algorithm~\ref{alg:ss1}  is the bandwidth $\kappa$. To show the performance of Algorithm~\ref{alg:ss1} for recovering the low-frequency components, we have the following results.

Denote $\Vm_{(\kappa)}$ be the first $\kappa$ columns of the inverse graph Fourier transform matrix $\Vm$, and $\Um_{(\kappa)}$ be the first $\kappa$ rows of the graph Fourier transform matrix $\Um$.
\begin{myLem} 
\label{lem:unbias_random}
Algorithm~\ref{alg:ss1} with bandwidth $\kappa$ provides an unbiased estimator of the first $\kappa$ frequency components, that is, 
\begin{equation*}
	\mathbb{E} \x^* \ = \ \Vm_{(\kappa)} \Um_{(\kappa)} \x,~~~{\rm for~all}~\x,
\end{equation*}
where $ \x^*$ is the result of Algorithm~\ref{alg:ss1}.
\end{myLem}
The advantage of Algorithm~\ref{alg:ss1} is its efficiency, that is, we only need the first $\kappa$ eigenvectors involved in the computation, which is appealing for large-scale graphs. The disadvantage is that when the main energy of an original graph signal is not concentrated in the first $\kappa$ frequency components, the recovered graph signal has a large bias.
\begin{myThm}
\label{thm:upper_ss1}
 For $\x \in \BLT(K, \beta, \mu)$, let $\x^*$ be the result of Algorithm~\ref{alg:ss1} with bandwidth $\kappa \geq K$, we have,
\begin{eqnarray*}
\mathbb{E}  \left\|  \x^* - \x  \right\| ^2 \ \leq \  \frac{ \alpha_2 \mu  \left\| \x \right\|_2^2 } {  \kappa^{2\beta} } +   \frac{ \alpha_2 ( \max_j x_j^2 + \sigma^2)    }{ |\M|} N \left\|  \Um_{(\kappa)} \right\|_F^2,
\end{eqnarray*}
where $\alpha_2$ is the stability constant of $\Vm$ in~\eqref{eq:GFTB}, $\sigma^2$ is the noise level in~\eqref{eq:sampling}, and $\left\|\cdot \right\|_F$ is the Frobenius norm.
\end{myThm}
Due to the limited space, we do not show the proof here. The main idea follows from the bias-variance tradeoff. The first term is the bias term, and the second terms is the variance term.  Since Algorithm~\ref{alg:ss1} can recover the first $\kappa$ frequency components on expectation, the bias comes from the other $(N-\kappa)$ frequency components, which can be bounded from the definition of~$\BLT(K, \mu, \beta)$ when $\kappa \geq K$. The variance term depends on $\left\|  \Um_{(\kappa)} \right\|_F^2$, which represents the graph structure.

\subsection{Recovery Strategy based on Experimentally Designed Sampling}
We consider the following recovery strategy.
\begin{myAlg}
\label{alg:ss2}
We sample a graph signal $|\M|$ times. Each time, we choose a node with probability $w_i = \left\| \u_i \right\|_2 / \sum_{j=0}^{N-1} \left\| \u_j \right\|_2$, where $\u_i$ is the $i$th column of $\Um_{(\kappa)}$, and take a measurement $y_i$. We then recover the original graph signal  by using the following two steps:
\begin{eqnarray*}
  \widehat{x}_k^* & = & \frac{1}{ |\M|} \sum_{i \in \M }  \frac{1}{w_i} \Um_{ki} y_i,
  \\
  x_i^* & = & \sum_{k < \kappa} \Vm_{ik} \widehat{x}_k^*.
\end{eqnarray*}
where $ x_i^*$ is the $i$th component of the recovered graph signal~$\x^*$.
\end{myAlg}
Similarly to Algorithm~\ref{alg:ss1}, Algorithm~\ref{alg:ss2} aims to estimate the first $\kappa$ frequency components, and reconstructs the original graph signal based on these graph frequency components. The difference comes from the normalization factor. In Algorithm~\ref{alg:ss1}, the contribution from each measurement is normalized by  a constant, the size of the graph; and in Algorithm~\ref{alg:ss2}, the contribution from each measurement is normalized based on the norm of the corresponding column in $\Um_{(\kappa)}$, called~\emph{sampling scores}.  Sampling scores are similar to leverage scores used in the matrix approximation~\cite{DrineasMMW:12}, where the goal is to evaluate the contribution from each column to approximating matrix. Note that leverage scores use the norm square, $ \left\| \u_i \right\|_2^2$, and we use the norm, $ \left\| \u_i \right\|_2$.  When we use the norm square as sampling scores, the performance is the same with the random sampling.

We can show  that Algorithm~\ref{alg:ss2} is also an unbiased estimator for recovering the low-frequency components.
\begin{myLem} 
\label{lem:unbias_exp}
Algorithm~\ref{alg:ss2} with bandwidth $\kappa$ provides an unbiased estimator of the first $\kappa$ frequency components, that is, 
\begin{equation*}
	\mathbb{E} \x^* \ = \ \Vm_{(\kappa)} \Um_{(\kappa)} \x,~~~{\rm for~all}~\x,
\end{equation*}
where $\x^*$ is the result of Algorithm~\ref{alg:ss2}.
\end{myLem}
\begin{myThm}
\label{thm:upper_ss2}
 For $\x \in \BLT_{\Adj}(K, \beta, \mu)$, let $\x^*$ be the result of Algorithm~\ref{alg:ss2} with bandwidth $\kappa \geq K$, we have,
\begin{eqnarray*}
 \mathbb{E} \left\|  \x^* - \x  \right\| ^2 \ \leq \ \frac{ \alpha_2 \mu \left\| \x \right\|_2^2 } {  \kappa^{2\beta} }  +   \frac{  \alpha_2 ( \max_j x_j^2 + \sigma^2)    }{  |\M| }    \left\|  \Um_{(\kappa)} \right\|_{2,1}^2.
\end{eqnarray*}
\end{myThm}
The main idea also follows from the bias-variance tradeoff. We see that Algorithms~\ref{alg:ss1} and~\ref{alg:ss2} have the same bias by recovering the first $\kappa$ frequency components on expectation. When each column of $\Um_{(\kappa)}$ has  roughly similar energy,   $N\left\|  \Um_{(\kappa)} \right\|_F^2$ and $\left\|  \Um_{(\kappa)} \right\|_{2,1}^2$ are similar. However, when the energy is concentrated on a few columns,  $N \left\|  \Um_{(\kappa)} \right\|_F^2$ is much larger than $\left\|  \Um_{(\kappa)} \right\|_{2,1}^2$, in other words, Algorithm~\ref{alg:ss2} has a significant advantage over Algorithm~\ref{alg:ss1} when the associated graph structure is irregular.
\vspace{-1mm}
\subsection{Convergence Rates}
\label{sec:2graphs}
To discriminate the proposed recovery strategies, we propose two types of graphs, and compare the convergence rates of Algorithms~\ref{alg:ss1} and~\ref{alg:ss2} for each type of these two.

\begin{defn}
  \label{df:regular}
A graph $\Adj \in \R^{N \times N}$ is~\emph{type-1},  when 
\begin{equation*}
| \Um_{i,j} | \ = \ O(N^{-1/2}), ~~~{\rm for~all}~ i,j = 0, 1, \cdots, N-1,
\end{equation*}
where $\Um$ are the graph Fourier transform matrix of $\Adj$.
\end{defn}

For a type-1 graph, each element in $\Um$ has roughly similar magnitudes, that is, the energy evenly spreads to each element in $\Um$.
Some examples are discrete-time graphs, discrete-space graphs, and unweighted circulant graphs~\cite{SandryhailaKP:11a}. Based on Theorems~\ref{thm:upper_ss1}, and~\ref{thm:upper_ss2}, we conclude as follows. 

\begin{myCorollary}
\label{cor:regular}
Let $\Adj \in \R^{N \times N}$ be a type-1 graph, for the class $\BLT_{\Adj }(K, \beta, \mu)$.
\begin{itemize}
\item Let $\x^*$ be the results given by Algorithm~\ref{alg:ss1} with the bandwidth $\kappa  \geq K$, we have
\begin{eqnarray*}
 \mathbb{E}  \left(  \left\| \x^* - \x \right\|_2^2 \right) \ \leq \  C N |\M| ^{-\frac{2\beta}{2\beta+1}},
\end{eqnarray*}
where $C > 0$, and the rate is achieved when $\kappa$ is in the order of $|\M|^{1/(2\beta+1)}$ and upper bounded by $N$;
\item Let $\x^*$ be the results given by Algorithm~\ref{alg:ss2} with the bandwidth $\kappa \geq K$, we have
\begin{eqnarray*}
 \mathbb{E} \left(  \left\| \x^* - \x \right\|_2^2 \right) \ \leq \ C N |\M|  ^{-\frac{2\beta}{2\beta+ 1}},
\end{eqnarray*}
where $C > 0$, and the rate is achieved when $\kappa$ is in the order of $|\M|^{1/(2\beta+1)}$ and upper bounded by $N$.
\end{itemize}
\end{myCorollary}
When $|\M| \gg N$, we set $\kappa = N$, and then the bias term is zero, and both upper bounds are actually $CN |\M|^{-1}$. We see that Algorithms~\ref{alg:ss1} and~\ref{alg:ss2} have the same convergence rate, that is, experimentally designed sampling does not perform better than random sampling for the type-1 graphs.

\begin{defn}
  \label{df:irregular}
 A graph $\Adj \in \R^{N \times N}$ is~\emph{type-2} with parameter $K_0 > 0$, when 
 $$
 \left\| \h^{(K)}_{T^c}  \right\|_1 \leq \alpha  \left\| \h^{(K)}_T  \right\|_1,~~~{\rm for~all}~K \geq K_0,
 $$
 where  $h^{(K)}_i =  \sqrt{ \sum_{k=0}^{K-1} \Um_{k,i}^2} $, $T$ indexes the largest $K$ elements in $\h$, $T^c$ indexes the other $(N-K)$ elements,  and $\alpha > 0$ is a constant.
 \end{defn}
A type-2 graph requires the sampling scores to be approximately sparse. When we take the first $K \geq K_0$ rows to form a submatrix, the energy in the submatrix concentrates in a few columns. 

Based on Theorems~\ref{thm:upper_ss1}, and~\ref{thm:upper_ss2}, we conclude the following.
\begin{myCorollary}
  \label{cor:irregular}
Let $\Adj \in \R^{N \times N}$ be a type-2 graph with parameter $K_0$, for the class $\BLT_{\Adj }(K, \beta, \mu)$.
\begin{itemize}
\item Let $\x^*$ be the results given by Algorithm~\ref{alg:ss1} with the bandwidth $\kappa \geq K$, we have
\begin{eqnarray*}
 \mathbb{E}  \left(  \left\| \x^* - \x \right\|_2^2 \right) \ \leq \  C N |\M|  ^{-\frac{2\beta}{2\beta+1}},
\end{eqnarray*}
where $C > 0$, and the rate is achieved when $\kappa$ is in the order of $|\M|^{1/(2\beta+1)}$ and upper bounded by $N$;

\item Let $\x^*$ be the results given by Algorithm~\ref{alg:ss2} with the bandwidth $\kappa \geq \max \{ K, K_0 \}$, we have
\begin{eqnarray*}
 \mathbb{E} \left(  \left\| \x^* - \x \right\|_2^2 \right) \ \leq \  C N |\M|^{-\frac{2\beta}{2\beta+2 - \gamma}} \ \leq \ C' N |\M|^{-\frac{2\beta}{2\beta+1}},
\end{eqnarray*}
where $C > 0$, the rate is achieved when $\kappa$ is in the order of $|\M|^{1/(2\beta+2-\gamma)}$ and upper bounded by $N$, and
\end{itemize}
\end{myCorollary}
\vspace{-6mm}

\begin{eqnarray*}
\gamma \in [ \max \{ 1, 2 \beta + 2 -  \frac{ \log |\M| }{ \log \max \{ K, K_0 \} } \},    
\\
 \max \{ 1, \frac{ {(2\beta+2) \log N } }{ {( \log N + \log |\M| )} } \} ].
\end{eqnarray*}

Similarly to the type 1 graphs, when $|\M| \gg N$, we set $\kappa = N$, and then the bias term is zero, and both upper bounds are $CN |\M|^{-1}$. We see that  Algorithm~\ref{alg:ss2} has a larger convergence rate  than Algorithm~\ref{alg:ss1}, that is, experimentally designed sampling exhibits much better performance than random sampling for the type-2 graph. The advantages follow from that, for type-2 graphs, $\left\|  \Um_{(\kappa)} \right\|_{2,1}^2$ is in the order of $\kappa^2$, and $N \left\|  \Um_{(\kappa)} \right\|_{F}^2$ is in the order of $N\kappa $.

We propose Definition~\ref{df:irregular} to obtain the asymptotic results, however, it is too strict to model real-world graphs. When considering a small sample size and a small bandwidth, we just need the requirement in Definition~\ref{df:irregular} holds for some $K = K_0 \leq N$. We call those graphs as the general type-2 graphs. Simulations shows that scale-free scales belong to the general type-2 graphs.

\vspace{-1mm}
\subsection{Relation to Graph Signal Inpainting}
Graph signal inpainting aims at recovering globally smooth graph signals based on random sampling~\cite{ChenSMK:14}. It solves the following optimization problem,
\vspace{-1mm}
\begin{subequations}
\label{eq:gsi}
  \begin{eqnarray}
   	\x^* &=& \arg \min_\x \,  \left\|  \x - \Adj \x \right\|_2^2, \\
   \label{eq:Opt_cond}
	&& \text{subject to} ~~\left\| \Psi \x - \y \right\|_2^2 ~\leq~ \sigma^2,
   \end{eqnarray}
 \end{subequations}
where $\sigma^2$ is noise level, $\y$ is a vector representation of the noisy measurements~\eqref{eq:sampling}, and $\Psi$ is the sampling operator~\eqref{eq:Psi}. Graph signal inpainting focuses on recovery in the vertex domain, and  the proposed recovery strategies  focus on  recovery in graph spectral domain. The optimum of~\eqref{eq:gsi} guarantees that the recovered graph signal is close to the measurements at given nodes, but Algorithms~\ref{alg:ss1} and~\ref{alg:ss2} guarantee the recovery of the low-frequency components.

\subsection{Relation to  Sampling Theory on Graphs}
Sampling theory on graphs aims at recovering bandlimited graph signals based on both the random sampling and the experimentally designed sampling~\cite{ChenVSK:15}. It solves the following optimization problem,
  \begin{eqnarray}
  \label{eq:st}
   	\x^* = \argmin_{\x \in \BL_{\Adj}(K) } \,  \left\|  \Psi \x -  \y \right\|_2^2 
=  \Vm_{(K)}  (\Psi  \Vm_{(K)}  )^{+} \y,
   \end{eqnarray}
where $\Psi$ is the sampling operator~\eqref{eq:Psi}, $\y$ is a vector representation of the noisy measurements~\eqref{eq:sampling}, and $(\cdot)^+$ is the pseudo-inverse. 
When the original graph signal is bandlimited, $\x \in \BL_{\Adj}(K)$, it is clear that the result of~\eqref{eq:st} is an unbiased estimator of $\x$. When the original graph signal is not bandlimited, the result of~\eqref{eq:st} is a biased estimator of the first $K$ frequency components, because the signal belonging to the other frequency band also projects onto the first $K$ components. In a sense of recovering the low-frequency components,~\eqref{eq:st} needs fewer samples, but Algorithms~\ref{alg:ss1} and~\ref{alg:ss2} are more reliable.

\begin{figure*} 
  \begin{center}
    \begin{tabular}{ccc}
\includegraphics[width=0.5\columnwidth]{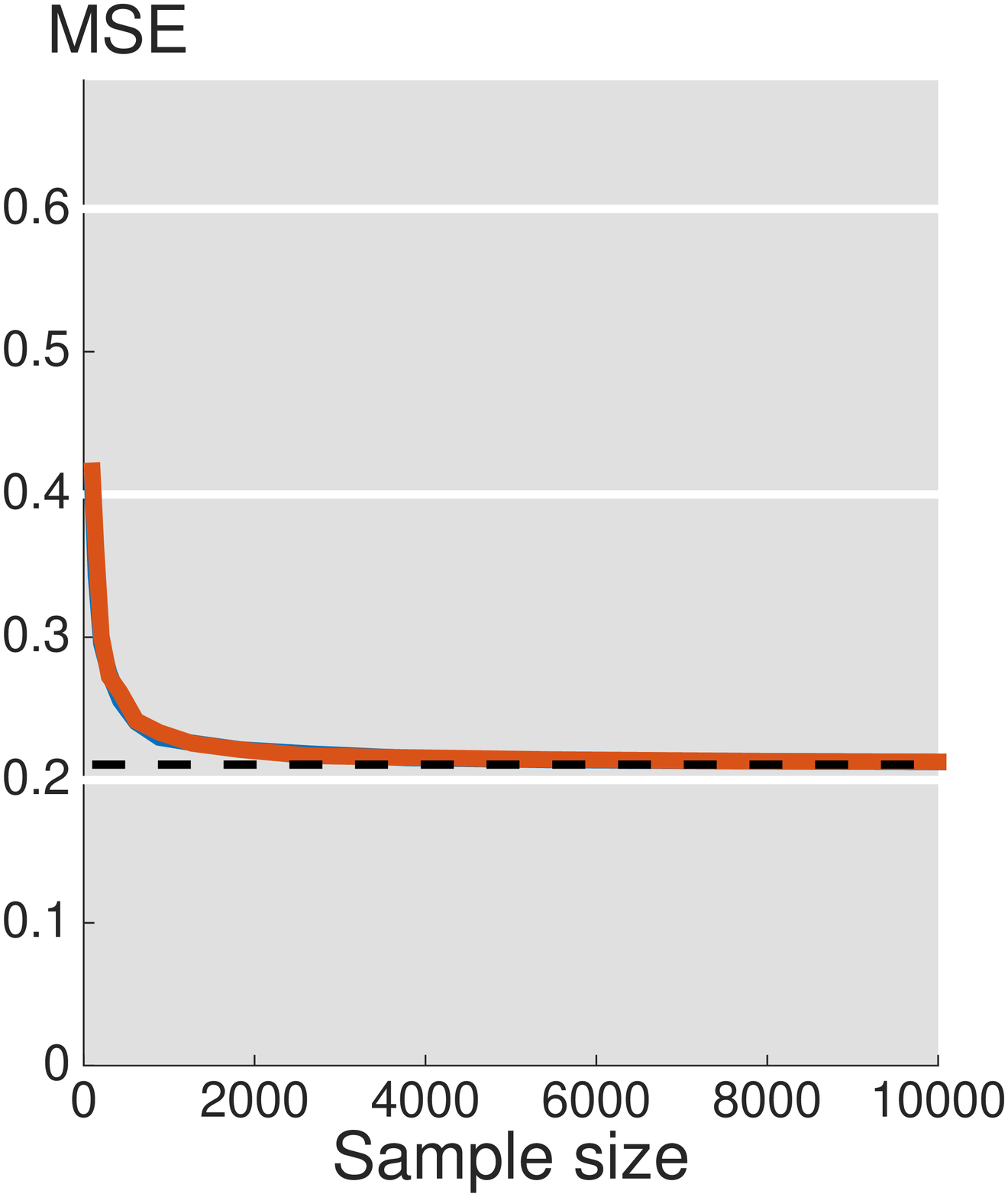}  & \includegraphics[width=0.5\columnwidth]{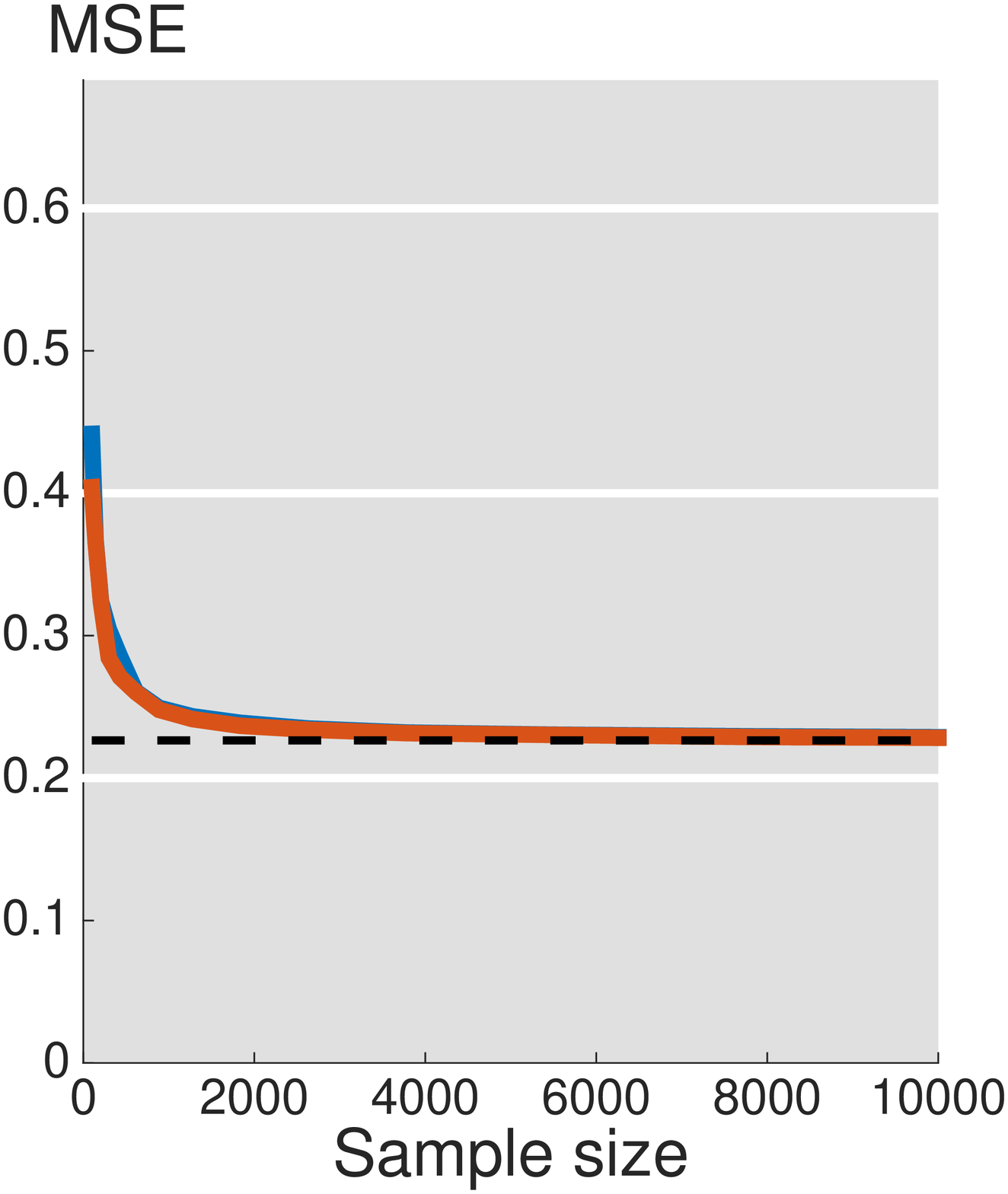}  &
\includegraphics[width=0.5\columnwidth]{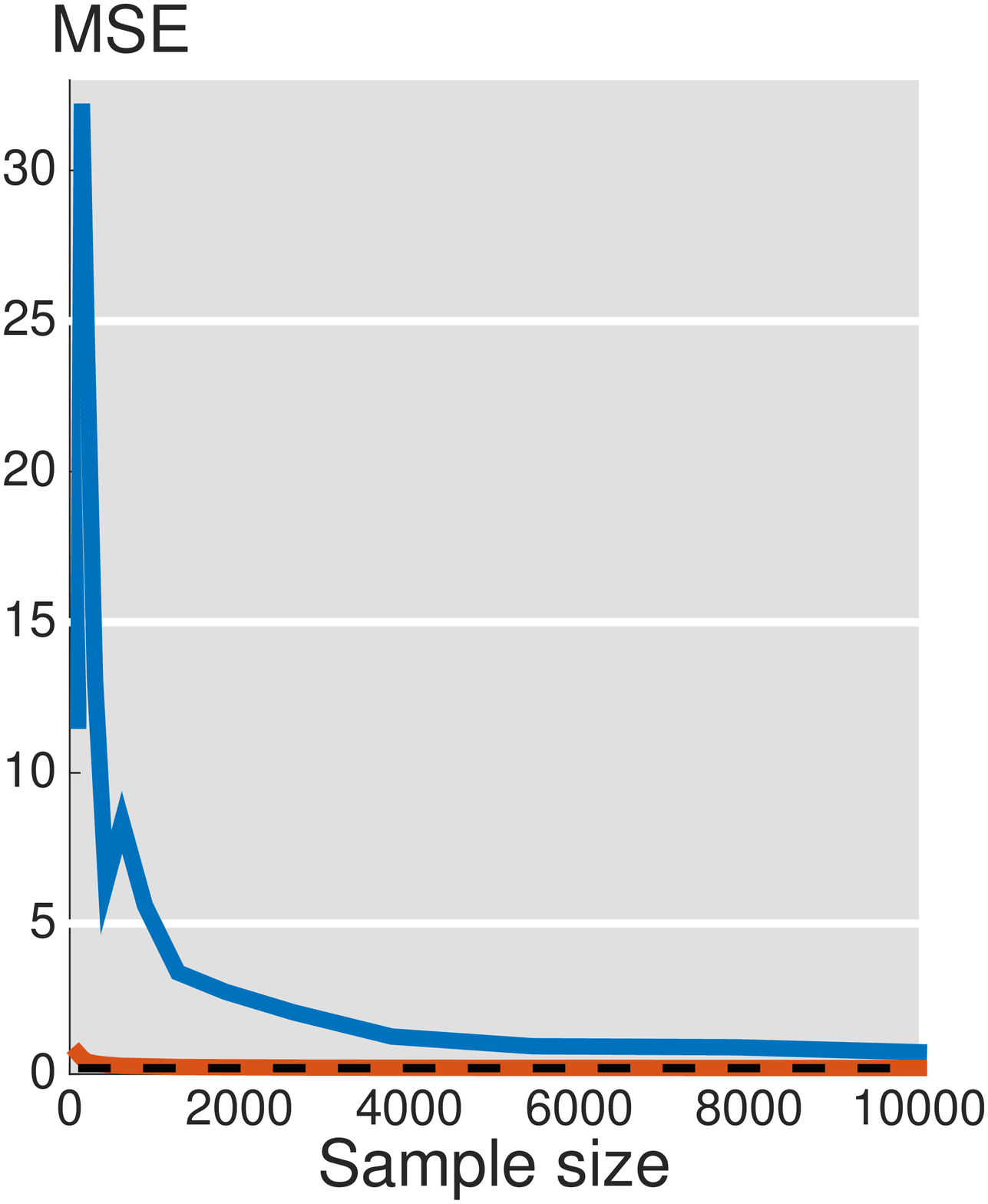}  \\
      {\small (a) Ring graph with 4-nearest neighbor .}   &  {\small (b) Erd\H{o}s-R\'enyi graph.}  &   {\small (c) Star graph.}      
    \end{tabular}
  \end{center}
  \caption{\label{fig:ring} Comparison of recovery error of Algorithm~\ref{alg:ss1} (blue curve) and~\ref{alg:ss2} (red curve). MSE indicates the mean square error. }   
\end{figure*}

\section{Experimental Results}
\label{sec:examples}
In this section, we compare the empirical performance of Algorithms~\ref{alg:ss1} and~\ref{alg:ss2} on three specific graphs:  a ring graph, an Erd\H{o}s-R\'enyi  graph, and a star graph.

For each graph $\Adj$, we generate 50 graph signals by the following two steps. We first generate the graph frequency components as
\begin{eqnarray*}
\label{eq:simulation}
\widehat{x}_k
\left\{
	\begin{array}{ll}
		 \sim \mathcal{N}(1, 0.5) & \mbox{if } k < K, \\
		 =  {K^{2\beta}}/{k^{2\beta}} & \mbox{if } k \geq K.
	\end{array}
\right.
\end{eqnarray*}
We then normalize $\xhat$ to have norm one, and obtain $\x = \Vm \xhat$. It is clear that $\x \in \BLT_{\Adj } (K, \beta, \mu)$, where $K = 10$ and $\beta = 1$. During the sampling, we simulate the noise $\epsilon \sim \mathcal{N}(0, 0.01) $. In the recovery, we set the bandwidth $\kappa$  to 10 for both algorithms. We consider the following three graphs.

\mypar{Ring Graph with $k$-nearest Neighbors}
We consider a graph with each node connecting to its $k$-nearest neighbors. The eigenvectors are similar to the discrete cosine transform and the energy evenly spreads to each element in $\Um$~\cite{SandryhailaKP:11a}, which follows Definition~\ref{df:regular}. Based on Corollary~\ref{cor:regular}, we expect that Algorithm~\ref{alg:ss2} has a similar performance with Algorithm~\ref{alg:ss1}. In the simulation, the ring graph has 10,000 nodes, and each node connects to its 4 nearest~neighbors. 

\mypar{Erd\H{o}s-R\'enyi Graph}
We consider a random graph where each pair of nodes is connected with some probability, also known as an Erd\H{o}s-R\'enyi graph~\cite{Newman:10}. Since the maximum value of eigenvectors of an Erd\H{o}s-R\'enyi graph is bounded by $O(N^{-1/2})$~\cite{TranVW:13}, the energy also spreads to each element in $\Vm$, which follows Definition~\ref{df:regular}. Based on Corollary~\ref{cor:regular},  we expect that Algorithm~\ref{alg:ss2} has a similar performance with Algorithm~\ref{alg:ss1}. In the simulation, the Erd\H{o}s-R\'enyi  graph has 10,000 nodes, and each pair of nodes is connected with probability of 0.01, that is, each node has 100 neighbors on expectation. 

\mypar{Star Graph}
We consider a graph with a central node connecting to all other nodes, known as the star graph.  The simulations show that star graphs approximately follows Definition~\ref{df:irregular}. Based on Corollary~\ref{cor:irregular}, we expect that Algorithm~\ref{alg:ss2} outperforms Algorithm~\ref{alg:ss1}. In the simulation, the star graph has 10,000 nodes.

\mypar{Results} 
Figure~\ref{fig:ring}  compares the performances between Algorithms~\ref{alg:ss1} and~\ref{alg:ss2} averaged over 50 tests. The blue curve represents Algorithm~\ref{alg:ss1}, the red curve represents Algorithm~\ref{alg:ss2}, and the black dotted line represented the linear approximation by the true first $K$ frequency components. We see that both algorithms converges to the linear approximation by the first $K$ frequency components, which supports the results in Lemmas~\ref{lem:unbias_random} and~\ref{lem:unbias_exp}. For two type-1 graphs, including the ring graph with 4-nearest neighbors and an Erd\H{o}s-R\'enyi graph, Algorithms~\ref{alg:ss1} and~\ref{alg:ss2} provide similar results; however, for the star graph, Algorithm~\ref{alg:ss2} performs much better than Algorithm~\ref{alg:ss1}, which supports the results in Corollaries~\ref{cor:regular} and~\ref{cor:irregular}.

\section{Conclusions}
\label{sec:conclusions}
 We proposed a new class of smooth graph signals, called approximately bandlimited, and we then proposed two recovery strategies based on random sampling and experimentally designed sampling. We showed that both strategies are unbiased estimators for the low-frequency components, and experimentally designed sampling outperforms random sampling when a graph is irregular. We validate the recovery strategies on three specific graphs: a ring graph, an Erd\H{o}s-R\'enyi  graph, and a star graph. The simulation results  support the theoretical analysis.

\bibliographystyle{IEEEbib}
\bibliography{bibl_jelena}
\end{document}